\documentstyle[11pt,paspconf,epsf]{article}

\begin{document}

\title{The Intermediate Line Region and the Baldwin Effect}
\author{M. S. Brotherton}
\affil{Institute of Geophysics and Planetary Physics, Lawrence Livermore National Laboratory}
\author{Paul J. Francis}
\affil{Mt Stromlo \& Siding Spring Observatory and Department of Physics and 
Theoretical Physics, Australian National University}

\begin{abstract}

Statistical investigations of samples of quasars have established
that clusters of properties are correlated.  The strongest trends among
the ultraviolet emission-line properties are characterized by the
object-to-object variation of emission from low-velocity gas, the so-called
``intermediate-line region'' or ILR.  The strongest trends among the optical 
emission-line properties are characterized by the
object-to-object variation of the line intensity ratio of [O III] $\lambda$5007
to optical Fe II.
Additionally, the strength of ILR emission correlates with [O III]/Fe II, 
as well as with radio and X-ray properties.  The fundamental physical parameter
driving these related correlations is not yet identified.
Because the variation in the ILR dominates the variation in the equivalent
widths of lines showing the Baldwin effect, it is important to 
understand whether the physical parameter underlying this variation
also drives the Baldwin effect or is a primary source of scatter in 
the Baldwin effect.  

\end{abstract}

\keywords{quasars, emission lines}

\section{Introduction}

The optical/ultraviolet spectra of quasars are similar over a wide range of 
luminosities and radio properties.  The spectra are characterized by strong
continuum emission, broad ($\Delta v >$ 2000 km s$^{-1}$)
emission lines arising from a broad-line region (BLR), and narrow emission 
lines ($\Delta v <$ 1000 km $^{-1}$) arising from a more extended narrow-line 
region (NLR).  Early photoionization models of the
BLR (e.g., Baldwin \& Netzer 1978; Kwan \& Krolik 1981)
showed that it was possible to reproduce typical BLR line ratios with
a single type of AGN cloud (standard model reviewed and critiqued by, e.g., 
Ferland 1986), but it is clear that the BLR is heterogeneous: (1) high and
low-ionization lines are underpredicted by the standard model 
(e.g., Netzer 1985), (2) lines show ionization-dependent velocity shifts 
(Gaskell 1982; Espey et al. 1989, 1994), (3) different lines show different
time lags in response to continuum changes (e.g., Korista et al. 1995),
(4) different lines can show dramatic profile differences 
(e.g., Netzer et al. 1994).
The fact that single-zone models work as well as they do can be attributed
to the powerful selection effects of ``locally optimally emitting clouds''
(Baldwin et al. 1995; Ferland this volume); many emission lines in the optical
and ultraviolet are preferentially emitted from clouds with a narrow range
of properties.

Even if beset by strong selection effects that dominate its emissions, 
the BLR is still an important probe of the spatially unresolvable sub-parsec 
environment of quasars.  Gas to fuel the presumed supermassive black hole
central engine must pass through the BLR, as must ISM and IGM-enriching
outflows originating in disk winds or jets.  
Statistical relationships among broad-line and other properties provide a 
means of investigating physical parameters, such as the black hole
mass and accretion rate, that underlie the appearance of quasars.
As the size of carefully selected quasar samples grows, as well as the 
quality of data available for such samples, likewise grows the need for
more sophisticated statistical techniques. 

One such multivariate technique that has become increasingly applied in AGN 
studies and other areas of astrophysics is principal component analysis or PCA
(e.g., Bernstein 1988).  
Technically, PCA is the eigenanalysis of the correlation matrix
of a set of input variables; the results are the eigenvectors (or principal
components) and their corresponding eigenvalues.  The eigenvectors can be
visualized as the directions in parameter space described by the elliptical
axes of the scatterplot of input variables, and the eigenvalues as a 
quantification of the amount of variance in the direction of these axes.  
Eigenvector 1 is then the direction in $n$-dimensional parameter space that 
accounts for the most variation
in the data set, and can include correlations among many variables.
When PCA is effective, the many input variables (typically measured properties
that would a priori seem unrelated) can be transformed into a few eigenvectors
that may be interpreted as the effect of the important underlying physical 
processes.

The discussion that follows describes eigenvector 1 correlations in the 
ultraviolet and optical spectra of quasars, how they are related to each 
other and other quasar properties, what physics underlies eigenvector 1,
and how all of this is related to the Baldwin effect. 
The primary source of variance in the ultraviolet spectra of quasars
involves the equivalent widths of the emission lines, which is one of the 
components of the Baldwin effect.  If eigenvector 1 is luminosity independent
(points in a direction orthogonal to luminosity), then the physical parameter
underlying eigenvector 1 is the source of scatter in the Baldwin effect.
If eigenvector 1 depends on luminosity (points in a direction 
parallel to luminosity), then the physical parameter underlying eigenvector 1 
helps create the Baldwin effect.  As will be discussed, current data sets 
provide contradictory evidence for which is the case.

\section{Ultraviolet Eigenvector 1: The Intermediate Line Region}

Investigations of luminous quasars' broad UV lines
identified strong correlations involving emission-line widths,
shifts, equivalent widths, and ratios
(Francis et al. 1992; Wills et al. 1993; Brotherton et al. 1994a, b).
A simple model developed to explain these trends approximates UV broad lines
as emission from two regions, an intermediate-line region (ILR), and a 
very-broad-line region (VBLR), together comprising the traditional BLR.

This decomposition is a simple, approximate explanation for the observation
that broader lined quasars have smaller equivalent widths and different line 
ratios when compared to narrower lined quasars.  
Figure 1 illustrates this using 
a composite of narrow-lined quasar spectra (2000 km s$^{-1} < $FWHM$_{CIV} <
3500$\ km s$^{-1}$) and a composite of broad-lined quasar spectra
(6000 km s$^{-1} < $FWHM$_{CIV} < 8000$\ km s$^{-1}$).   
The difference spectrum, or ILR spectrum, produced this way is essentially 
identical to the first principal component (PC1) spectrum produced by
the spectral PCA of the Large Bright Quasar Survey (Brotherton et al. 1994b;
Francis et al. 1992).

\begin{figure}
\plottwo{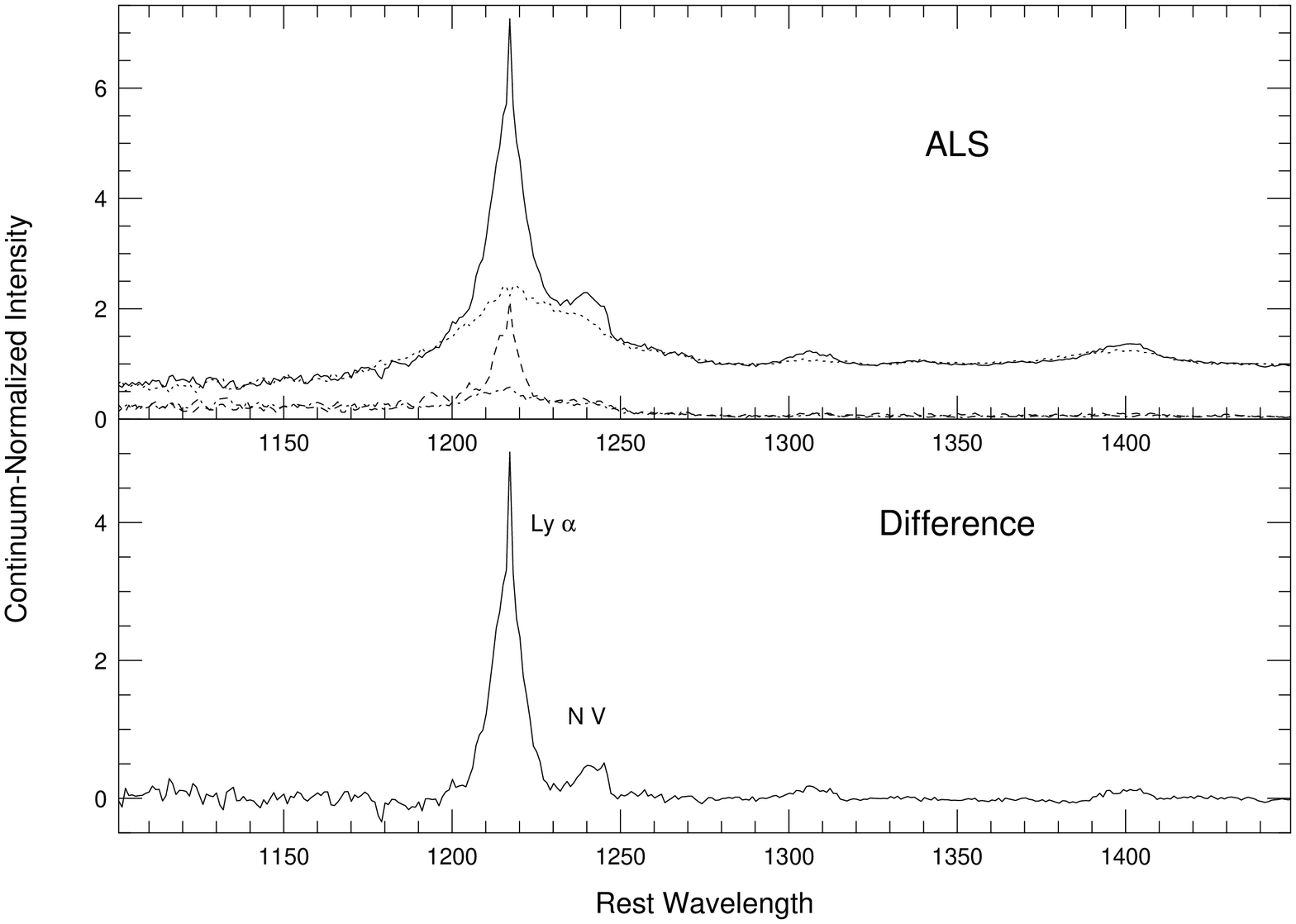}{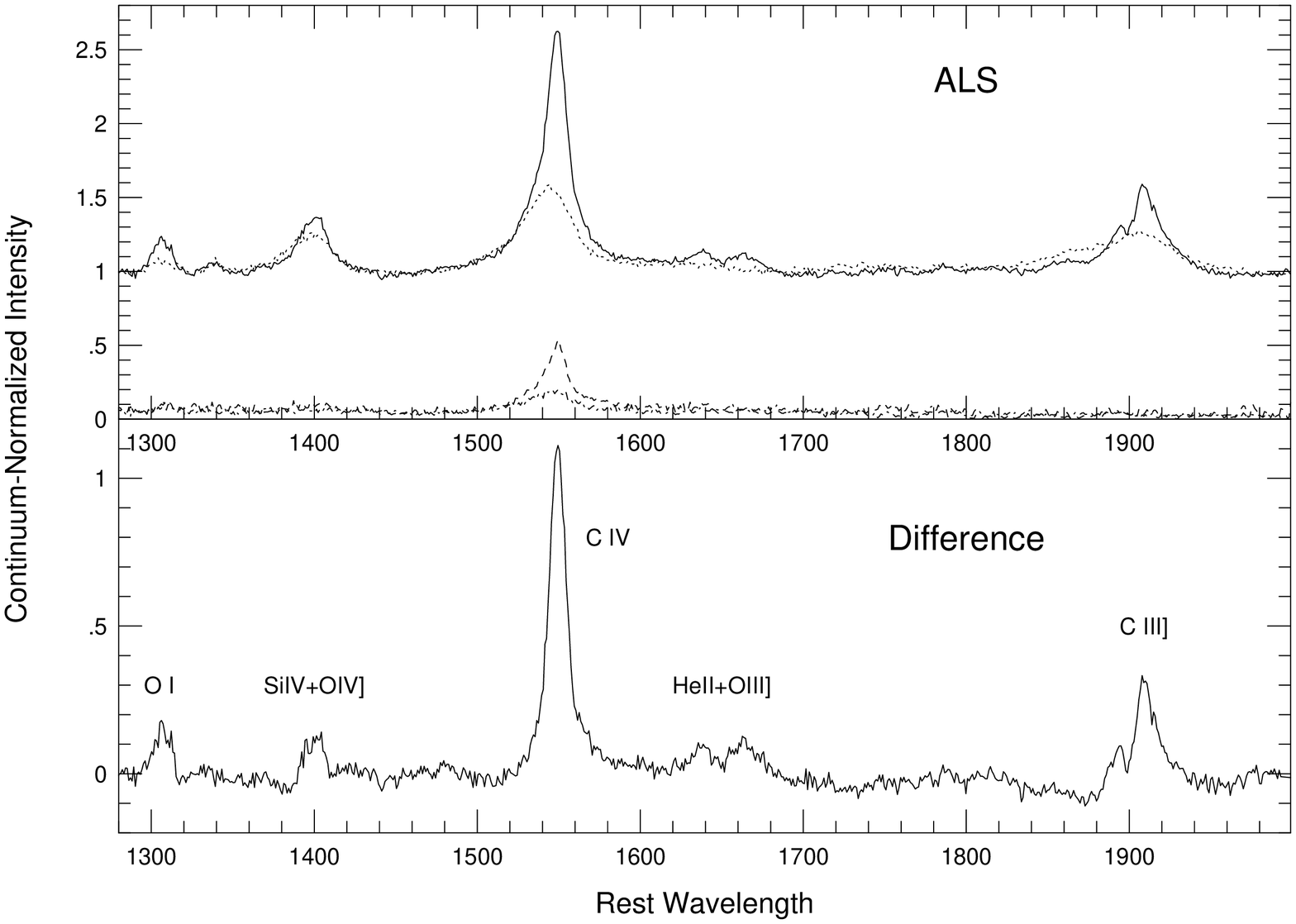}
\caption{The narrow ($solid$) and broad-lined ($dot\-ted$) 
continuum-normalized (``EW'') composite spectra of the Ly $\alpha$ region 
(left) and C IV (right).  The difference spectra are displayed below. 
From Brotherton et al. 1994b.}
\end{figure}

\subsection{Photoionization Modeling}

The emission-line ratios of the ILR (difference) spectrum and the VBLR
(broad-lined composite) can be modeled using photoionization codes
such as CLOUDY (Ferland 1993).  While single-zone models fail to reproduce
the highest and lowest ionization lines, a two-zone model does a 
better job reproducing the heterogeneous BLR.  Brotherton et al. (1994b)
showed that the ILR spectrum could be well modeled, whereas the VBLR spectrum
is probably still too heterogeneous for a good single-zone model.
The discriminating diagnostic lines are O III] $\lambda$1663, a
semi-forbidden line, and Al III $\lambda$1860, an important coolant in
high density clouds, which suggest that the ILR is more distant from the
nucleus and less dense than the rest of the BLR.  
The observed and derived properties of the ILR, VBLR, and the NLR are
tabulated (from Brotherton et al. 1994b).

\begin{table}
\begin{center}
\centerline{Table 1. Comparison of Emission-Line Regions}
\begin{tabular}{llll}
\tableline\tableline
Property & NLR & ILR & VBLR \\
\tableline
Velocity Dispersion (km s$^{-1}$) & $\sim$500& $\sim$2000& $\sim$7000 \\
Radial Distance (pc) & 10$^{2-3}$ & $\sim$1 & $\sim$0.1 \\
Gas Density ($n_H$, cm$^{-3}$) & 10$^{4-6}$ & $\sim$10$^{10}$ & $\sim$10$^{12.5} $ \\
Ionization Parameter (U = $\phi_i$/n$_H$) & $\sim$0.01 & $\sim$0.01 & $\sim$0.01 \\
Redshift cf. Systemic (km s$^{-1}$) & 0 & $\sim$0 & $\sim$ $-$1000 \\
Covering Factor ($\Omega$/4$\pi$) & $\leq$0.02 & $\leq$0.03 & $\sim$0.24 \\
\tableline
\end{tabular}
\end{center}
\normalsize
\end{table}

Keep in mind that these results were obtained by modeling spectra 
derived from the most luminous quasars known, and at least the
size scales can be expected to vary with luminosity (e.g., Kaspi et al. 1996).
 
This two-component BLR breakdown may be generalized to $n$-components as strong 
selection effects permit an ensemble of clouds experiencing a very wide range 
of physical conditions to reproduce, not badly, a typical quasar spectrum 
(Baldwin et al. 1995). The designations of ``ILR'' and ``VBLR,'' and more 
specifically the ratio of ILR to VBLR emission, may simply represent the limits
of such an ensemble distribution.  Differences in the relative emission of these
limits account for much of the diversity of broad-line profiles, as well
as relations among line strength, line width, asymmetry and peak blueshift.

Comparison with other AGN emission-line regions shows
that the ILR spectrum tends to be intermediate between that of the VBLR and 
that of gas more distant from the ionizing continuum, such as the NLR and 
extended Ly~$\alpha$ nebulosity.  This suggests that the ILR may be more 
properly identified as an inner extension of the NLR rather than as a 
component of the BLR, a hypothesis strengthened below (\S\ 4).

\section{Optical Eigenvector 1: The Fe II -- [O III] Anti-Correlation}

The object-to-object variation in the optical spectra of low-redshift
quasars is dominated by the inverse correlation between narrow
[O III] $\lambda$5007 (FWHM $\sim$ 500 km s$^{-1}$)
and optical Fe II emission (eigenvector 1 of the PCA of Boroson \& Green 1992
of the parameterized spectra of optically selected quasars from the Bright 
Quasar Survey, or BQS).   Figure 2 illustrates this trend.
It is important to note that it is not just the equivalent width (EW) of 
[O III] $\lambda$5007 involved in the correlation, but also the
luminosity of [O III] $\lambda$5007.  There are a number of secondary
properties also correlated with eigenvector 1:
quasars with prominent [O III] $\lambda$5007 and weak optical Fe II
preferentially have broad, red-asymmetric H$\beta$ and are radio-loud
and strong in hard (2 keV) X-rays (e.g., Corbin 1993).
Furthermore, Laor et al. (1997), using a complete subset
of the PG quasars, found that the soft X-ray spectral slope, $\alpha_x$,
is strongly correlated with [O III]/Fe II in the sense that strong Fe II
emitting objects have steep soft X-ray spectra (the extreme of these are
identified with the narrow-line Seyfert 1 objects).

\begin{figure}
\plotfiddle{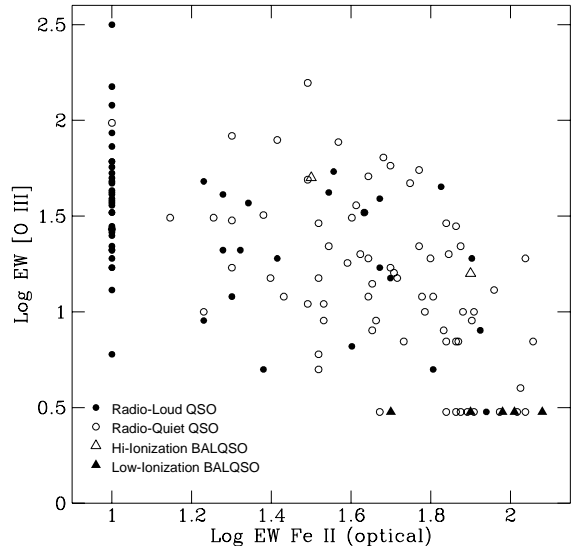}{6.6cm}{0}{39}{39}{-120}{-75}
\caption{Low-redshift quasars, after Fig. 2 of Wills \& Brotherton (1996).
EWs are restframe \AA.  Log EW[FeII] = 1 are $\sim$1.5$\sigma$ upper limits.
Log EW[OIII] = 0.5 are $\sim$3$\sigma$ upper limits.  Low-ionization BALQSOs
show excessive Fe II and negligible [O III] $\lambda$5007 emission.}
\end{figure}

\section{The ``Unified'' Eigenvector 1}

In order to study simultaneously the statistical behavior of both
optical and ultraviolet spectral properties, it is necessary to observe
a wide range of wavelengths: for low-redshift quasars, the optical and
the ultraviolet; for high-redshift quasars, the optical and
the near-infrared.  This has only recently become possible with the
same data quality as in the optical because of the Hubble Space Telescope (HST)
and new generations of near-IR detectors.

Brotherton (1996b) obtained near-IR spectra of the H$\beta$--[O~III] 
$\lambda$5007 region for 32 intermediate to high redshift quasars with 
a range in ILR strengths.  The strength of narrow-line emission, characterized
by [O III] $\lambda$5007 relative to the continuum and H$\beta$,
is indeed correlated with that of the line cores\footnote{The term ``line 
core'' refers to the ILR contribution alone, not simply the emission within
some velocity interval of the peak.} of C IV and C III], and 
inversely correlated with optical Fe II emission.
Eigenvector 1 in the optical and the ultraviolet is the same.
This result is corroborated by Wills et al. (this volume), who obtained
HST ultraviolet spectra of a subsample of the BQS.  

Marziani et al. (1996) and Wang et al. (1996--based on IUE data)
find that the strength of optical Fe II multiplets is inversely related
to the equivalent width of C IV $\lambda$1549.  This is consistent
with our result that the ILR emission (which is the main determinant of
EW C IV, Wills et al. 1993), is inversely correlated with optical Fe II
emission.  Thus the relationships among ILR, NLR, and the Fe II emission
appear to hold in lower redshift, lower luminosity quasars.
Table 2 summarizes a large but not exhaustive set of correlated
properties that together comprise a ``unified'' eigenvector 1.
If these quantities related by eigenvector 1 can be understood in terms
of the underlying physics, their variance might constitute a
``fundamental plane'' for quasars by analogy with the ``fundamental plane'' for
galaxies.  Therefore understanding eigenvector 1 may allow important physical
parameters to be estimated on the basis of a few easy-to-measure observables.

\begin{table}
\begin{center}
\centerline{Table 2. Correlated Eigenvector 1 Properties}
\begin{tabular}{lll}
\tableline\tableline
{Weak ILR} & {Strong ILR} & Ref. \\
\tableline
Broad Ly $\alpha$, C IV, C III] & Narrow Ly $\alpha$, C IV, C III] & 1, 2, 3, 4 \\
Small EW C IV & Large EW C IV & 1, 2\\
Small EW Ly $\alpha$ & Large EW Ly $\alpha$ & 1, 4 \\
Small Ly $\alpha$/C IV & Large Ly $\alpha$/C IV & 1, 4 \\
Large C IV/$\lambda$1400 Feature & Small C IV/$\lambda$1400 Feature & 2 \\
``Flat-topped'' C IV & ``Sharply Peaked'' C IV & 1, 2 \\
C IV and C III] Blueshifted & C IV and C III] at Systemic $z$ & 3 \\
Weak [O III] $\lambda$5007 & Strong [O III] $\lambda$5007 & 5, 6 \\
Strong Optical Fe II & Weak Optical Fe II & 5, 6 \\
Weak Radio-jets (Radio-quiet) & Strong Radio-jets (Radio-loud) & 3, 5, 6, 7 \\
Steep Soft-X-ray Slope & Flat Soft-X-ray-Slope & 8 \\
Small Hard X-ray Luminosity & Large Hard X-ray Luminosity & 6, 8, 9, 10 \\
Small [O III] $\lambda$5007 Luminosity & Large [O III] $\lambda$5007 Luminosity & 6 \\
Narrow H$\beta$ with Blue Wing & Broad H$\beta$ with Red Wings & 6, 8 \\
Mg II BALQSOs & no Mg II BALQSOs & 11, 12 \\
\tableline
\end{tabular}
\end{center}
REF. 1=Francis et al. 1992. 2=Wills et al. 1993.  3=Brotherton et al. 1994a.  4=Brotherton et al. 1994b.  5=Brotherton 1996b. 6=Boroson \& Green 1992. 7=Francis et al. 1993.  8=Laor et al. 1994, 1997.  9=Corbin 1993.  10=Green 1998.  11=Boroson \& Meyers 1992.  12=Wills \& Brotherton 1996.
\normalsize
\end{table}


\subsection{Physical Explanations}

There is as yet no generally accepted cause for the eigenvector 1 variance.
The large number of related properties poses a special problem as well as an 
opportunity.  Simple explanations are likely to fail because these 
properties represent conditions on all size scales associated with the AGN 
phenomenon.   For instance, while the soft X-ray slope and ionizing continuum
correlate with eigenvector 1 (e.g., Laor et al. 1997), this alone cannot
explain the extreme range in [O III] $\lambda$5007 equivalent width,
although it might explain some of the line ratio differences (e.g., Mushotsky 
\& Ferland 1984; Korista this volume).
While radio-loudness correlates with eigenvector 1, the trends appear to hold 
for radio-quiet samples alone, so this property is unlikely to be 
fundamental.  Rather we are faced with developing an explanation for some 
aspects of eigenvector 1 directly, and others more indirectly.
A few possibilities include:

\subsubsection{Orientation.}
The framework of orientation can explain the variation of
many of the eigenvector 1 properties, and probably contributes to the observed 
correlations.  The line intensity ratio of optical Fe II to [O III] 
$\lambda$5007 is larger in core-dominant quasars than in lobe-dominant quasars 
(e.g., Zheng \& O'Brien 1990; Jackson \& Browne 1991; 
Brotherton 1996a), radio-loud classes believed to differ because of their 
orientation to our line of sight (e.g., Orr \& Browne 1982; 
Wills \& Brotherton 1995).  The line width and asymmetry of H$\beta$ also
vary with core dominance (Wills \& Browne 1986; Brotherton 1996a).  Often these
trends have been explained in terms of anisotropic line and axisymmetric 
continuum emission (Jackson et al. 1989; Jackson \& Browne 1991).  
Others have invoked accretion 
disks as the source of the strong, possibly anisotropic Fe II emission (e.g.,
Collin-Souffrin, Hameury, \& Joly 1988; Kwan et al. 1995).
Hard X-ray emission may also vary consistently with inclination
(``face-on'' implies smaller 2 keV flux) if the hard X-rays are produced
by Comptonization by nonthermal electrons above an accretion disk
(Ghisellini et al. 1991).

Orientation appears to fall short in accounting for at least one key item: 
[O III] $\lambda$5007 luminosity.  Boroson \& Green (1992)
argued against orientation because the luminosity of [O III] $\lambda$5007,
which they took to be an isotropic property (Jackson et al. 1989),
correlated with eigenvector 1, and this was inconsistent with
the strong correlation between continuum luminosity (enhanced for ``face-on''
quasars in the beaming model, thus decreasing EW [O III] $\lambda$5007)
and [O III] $\lambda$5007 luminosity. 

An extrapolation of the results of Baker (1997) may provide a boost
for orientation, however.  Baker finds, for a large sample of
low-radio-frequency-selected quasars, evidence for aspect-dependent
extinction from dust toroidally distributed between
the BLR and NLR.  Trends of the Balmer decrement and optical slope
with core dominance are consistent with this interpretation.
At large angles, [O III] $\lambda$5007 emitting clouds may be partially 
obscured.  But it is not clear if the obscuration is enough to explain the 
range in observed luminosity.

Orientation may be involved in driving eigenvector 1, but if so, it requires 
several elements including beaming effects, dust reddening, and selection 
biases.

\subsubsection{Eddington Fraction.}
The Eddington fraction is the ratio between the luminosity of an
accreting mass and its Eddington luminosity (the point at which radiation
pressure balances gravity for accreting material).  Laor et al. (1997), 
following the suggestion that steep $\alpha_x$ quasars are analogous to 
`high'-state Galactic black hole candidates (e.g., White, Fabian, \& Mushotsky
1984; Pounds, Done, \& Osborne 1995), explained the $\alpha_x$ vs. 
FWHM H$\beta$ correlation in terms of range of Eddington fraction: for a 
given luminosity ``narrow'' broad lines (i.e., H$\beta$)
imply a higher Eddington fraction if the line width is gravitational.
An additional point in favor of this interpretation is that a
steep soft $\alpha_x$ is predicted to arise from a weaker hard X-ray component,
and for the $ROSAT$-observed sample of Laor et al. (1997), it does
appear to be changes in the hard X-rays leading to changes in
$\alpha_x$.  Also see Wandel \& Boller (1998).

Boroson \& Green (1992) also argued that the Eddington fraction was
the important parameter.  They surmised that optical Fe II emission was 
dependent on the covering fraction of the BLR, and that more BLR clouds 
(and hence higher accretion rate) would obscure the more distant NLR. 
Thus the covering fraction increases from the radio-loud strong [O III] 
$\lambda$5007, weak Fe II quasars to the radio-quiet weak 
[O III] $\lambda$5007, strong Fe II quasars.  They also noted that PG 
1700+518, a broad absorption line quasar or BALQSO, is found at the high 
covering fraction end.

\subsubsection{Age.}

This explanation is related to the above in the details, but is more 
fundamental.  Sanders et al. (1988) proposed a scenario in which galaxy mergers
produce dust-rich ultraluminous infrared galaxies, which then evolve
into quasars as the dust and gas are blown out by the AGN activity.
This fueling episode might correspond to a high Eddington fraction.

BALQSOs such as PG 1700+518 might then be characterized as young
or recently refueled quasars. Boroson \& Meyers (1992) noted that 
low-ionization BALQSOs constitute 10\% of IR-selected quasars 
(not the 1\% of optically selected quasars), and that they show very
weak narrow [O III] $\lambda$5007 emission, and, in an HST survey, Turnshek et
al. (1997) find that 1/3 of weak [O III] $\lambda$5007 quasar show BALs.
Voit et al. (1993) argue that low-ionization BALs are a manifestation of a 
``quasar's efforts to expel a thick shroud of gas and dust.''  {\em All} of the
low-ionization BALQSOs in Figure 2 lie at the extreme low [O III] $\lambda$5007,
high Fe II corner.
PG 1700+518 shows evidence for a recent interaction: a nuclear starburst ring
(Hines et al. 1998) and a companion galaxy with a 100 million year old
starburst (Stockton et al. 1998).  

Such an environment with a high covering factor of high density/column density
clouds can quench or frustrate radio jets (at least for certain models of 
jets), which would explain the anti-correlation between the presence of BALs 
and radio power (Stocke et al. 1992; Brotherton et al. 1998) and the
association of radio-loud quasars with strong NLR and ILR emission (Fig. 2;
Francis, Hooper, \& Impey 1993).

\section{Application to the Baldwin Effect}

Even without a physical explanation for eigenvector 1, the relationships
can be used empirically.
In the investigations of Wills et al. (1993) and Brotherton et al. (1994a),
C IV $\lambda$1549 did not display a significant Baldwin effect.  These samples
covered only a small range in quasar luminosity.  The EW$_{CIV}$ 
strongly correlated with FWHM$_{CIV}$, suggesting eigenvector 1 to be 
the source of scatter in the Baldwin effect.  Multiple regression using
both EW$_{CIV}$ {\em and} FWHM$_{CIV}$ as predictors of luminosity should
produce a tightened Baldwin effect if this hypothesis is true.

The Large Bright Quasar Survey or LBQS (Hewett et al. 1995) is the largest
complete optically selected sample anyone has yet studied in detail 
(although the luminosity range is relatively small).
Francis et al. (1992) measured the Baldwin effect for a 
high-redshift subsample of the LBQS, both for the entire C IV $\lambda$1549
line and also the line core and line wings separately.  They found marginally 
significant differences suggesting that the line cores contributed most
to the effect.

Figure 3 shows data from the LBQS investigation by Francis et al. (1992),
with an addition: a vector showing the magnitude and direction of eigenvector 
1 (from their spectral PCA) for each quasar. The distribution does not 
appear independent of luminosity: the left part of
the plot is heavy with up vectors, the right side with down vectors.
Eigenvector 1 is a primary cause of the Baldwin effect in this sample.
Correcting for eigenvector 1 would not reduce the Baldwin effect scatter, 
but the Baldwin effect itself. 

\begin{figure}
\plotfiddle{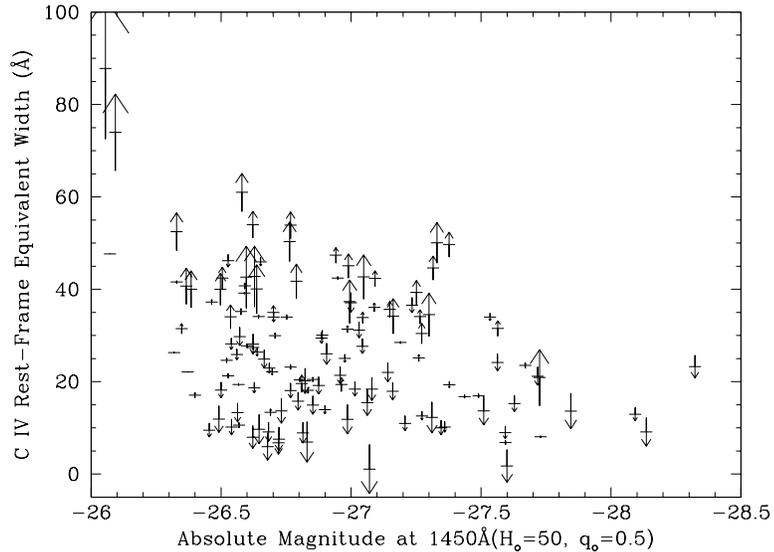}{6.5cm}{-90}{38}{38}{-150}{220}
\caption{The Baldwin effect in the LBQS sample from Francis et al. (1992).
The arrows on each point indicate the value of principal component 1 (PC1) from
their spectral principal component analysis.  Large up arrows indicate 
narrow peaky C IV lines, and large down arrows indicate broad, flat-topped
profiles.  The fact that the distribution of PC1 weights changes with 
luminosity suggests that its variation also drives the Baldwin effect.} \label{fig-1}
\end{figure}

Osmer, Porter, \& Green (1994) created composite spectra for samples of
quasars with different luminosities.  Difference spectra showed that the
change in emission-line equivalent width was confined to the low-velocity
gas.  



Unfortunately the situation appears to be different at low luminosities.
Boroson \& Green (1992) identify luminosity with eigenvector 2 in their
PCA of the BQS.  Wills et al. (this volume) also identifies luminosity and
the Baldwin effect with an eigenvector 2.  The luminosity ranges of these 
samples are again not ideal for investigating the Baldwin effect.


\section{Conclusions and Future Directions}

There is conflicting evidence about whether or not eigenvector 1 correlations
are driving the Baldwin effect or driving scatter in the Baldwin effect.
The answer is of course high-quality data on a large carefully selected sample
covering a wide range of luminosity, followed by multivariate analysis.  
An appropriate data set does not as yet appear to exist.

\acknowledgments

I would like to thank Bev Wills for her contributions,
both tangible and intangible.
This work has been performed under the auspices of the U.S. Department of Energy
by Lawrence Livermore National Laboratory under Contract W-7405-ENG-48.

\end{document}